# Development of a Parallel-Plate Avalanche Counter with Optical Readout (O-PPAC)

**Marco Cortesi,**[a,*] **Yassid Ayyad**[a], **and John Yurkon**[a]

[a] *National Superconducting Cyclotron Laboratory (Michigan State University),*
*East Lansing, Michigan 48824, U.S.A.*
E-mail*: cortesi@nscl.msu.edu*

ABSTRACT: we describe a novel gaseous detector concept for heavy-ion tracking and imaging: the Optical Parallel-Plate Avalanche Counter (O-PPAC). The detector consists of two thin parallel-plate electrodes separated by a small (typically 3 mm) gap filled with low-pressure scintillating gas (i.e. $CF_4$). The localization of the impinging particles is achieved by recording the secondary scintillation, created during the avalanche processes within the gas gap, with dedicated position-sensitive optical readouts. The latter may comprise arrays of collimated photo-sensors (e.g. SiPMs) that surround the PPAC effective area. We present a systematic Monte Carlo simulation study used to optimize the geometry of the OPPAC components, including SiPMs effective area, collimator dimensions, and operational conditions. It was found that the optimal design for 10x10 $cm^2$ OPPAC detector comprises four arrays, each of them counting a total of 15-20 individual photo-sensors. This configuration provides a localization capability with a resolution below 1 mm and good response uniformity. An experimental investigation successfully demonstrated the proof of principle stage of an O-PPAC prototype equipped with a single array of 10 photo-sensors, separated by 6 mm. The performance of the prototype was investigated with an LED light, under $^{10,12}C$ beam irradiation, and with a low-intensity 241-Am alpha-particle source. The experimental data obtained with the prototype is compared to the results obtained by systematic Monte Carlo simulations.

---

[*] Corresponding author.

**Contents**

## 1. Introduction

More than a century after the discovery of the basic principles of the Townsend avalanche process in gas media, gaseous detectors are still fundamental components at the frontier of present and next-generation particle/nuclear physics experiments.

For instance, heavy ion accelerator facilities requires efficient and spatially sensitive gaseous counters to monitor and track the beam particles along the beam line and at the target station, for particle-identification (PID) and momentum vector reconstruction on an event-by-event basis. Their advantages compared to solid-state detector technologies include low material budget and small thickness that lead to full transmission efficiency with minimal straggling, low charge-state heavy ion production, small beam-induced background and lower channeling effects that results in a more efficient PID.

The localization of the impinging charged particle in a conventional proportional counter (PC) is determined from the amplitudes of the signals recorded by segmented readout electrodes. In the two-dimensional Parallel-Plate Avalanche Counters – PPACs [1]–[4], the charge avalanche signals are readout by two orthogonal striped foils connected to a resistive divider chain, on either side of a central biased electrode. The four signals at the ends of the two chains are amplified, shaped and the peak voltages recorded. The X and Y positions are inferred from the ratio of the pulse-height amplitudes measured at each end of the resistor chains [5]. Alternatively, the electrode strips may be connected to multi-tapped delay-lines, and the position is determined from the difference in arrival times of the event signals at the two ends of the delay-lines [6].

The spatial resolution of the PPACs depends on several factors, including charge readout methods (charge division or delay-line), the dimension of the segmented readout, the detector gas gain, the amount of primaries released by the impinging beam particle in the detector effective volume, and on the granularity of the readout [6]. Strips with a sub-millimeter gap and a center-to-center separation below 1 mm are difficult to realize, limiting the spatial resolutions of the PPACs to 1 mm or above (FWHM). In addition, the counting rate capability of conventional PPACs with charge-division readout methods is limited to a few tens of KHz, while the delay-line PPAC is of a few hundred of KHz [7]. However, delay-line PPAC have generally lower detection efficiency due to a worse signal-to-noise ratio compared to charge-division based devices.

On the other hand, electroluminescence-based detectors (EL) are emerging as alternative solutions to conventional PC based on charge readout, in rapid developments thanks to the introduction of novel, high-sensitivity solid-state photo-detector technologies [8]–[12]. The operational principle of position-sensitive EL detectors is based on reading out the scintillation light produced via secondary mechanisms in the course of the electron avalanche process. The position of the impinging particles is computed as the center of the gravity of the light recorded by position-sensitive photo-detectors.

The wide range of applications that rely on the most advanced EL detectors includes double-phase liquid noble gas time projection chamber (TPC) for rare event searches (e.g. dark matter [8], [13]–[17], neutrino physics [18]–[21], low energy X-rays [22], [23], and nuclear medical imaging [24]. New technologies such as linear (APD) or Geiger (SiPM) Avalanche Photo-Diode [25], hybrid devices [26] and fast Charge-Couple Device CCD [27], have extremely low sensitivity and allow EL-based readout to deliver better signal-to-noise ratio and better energy resolution compared to the traditional charge-based detectors. Devices such as SiPMs and APDs are compact and may be produced of various sizes, which enables high versatility in designing high-granularity readout configurations.

In the present work we report the development of a novel gaseous detector concept based on EL readout, for heavy-ion tracking and imaging: the Optical Parallel-Plate Avalanche Counter (O-PPAC). We discuss the operational principle and present a comprehensive study on main expected performance and geometry optimization, carried out by systematic Monte Carlo simulations. We also report on the experimental investigation of a first O-PPAC prototype, consisting of a single position-sensitive array of SiPMs, tested under various irradiation conditions. Experimental data are compared to the Monte Carlo simulation results.

## 2. The O-PPAC

### 2.1 The operational principle

The basic design of an O-PPAC comprises two parallel, conductive electrodes, separated by a small gap (typically 3 mm wide). The gap is filled with a low pressure scintillating gas mixture, characterized by high electroluminescence light yield. Arrays of small, collimated photo-sensors (e.g. SiPMs) are arranged along the edges of the avalanche gap. When an ionizing particle crosses the detector active volume, it releases a small amount of energy in the form of ionization electrons along its track. These electrons are multiplied in the gas by the action of the uniform electric field established between the two parallel plates. The scintillation light, emitted during the avalanche processes by inelastic collisional excitations, is recorded by the arrays of collimated photo-sensors (Figure 1). The collimation is important for it narrows down the detected electroluminescence light distribution, such that its peak is weighted more heavily near the position of the avalanche, allowing precise localization of the impinging particles.

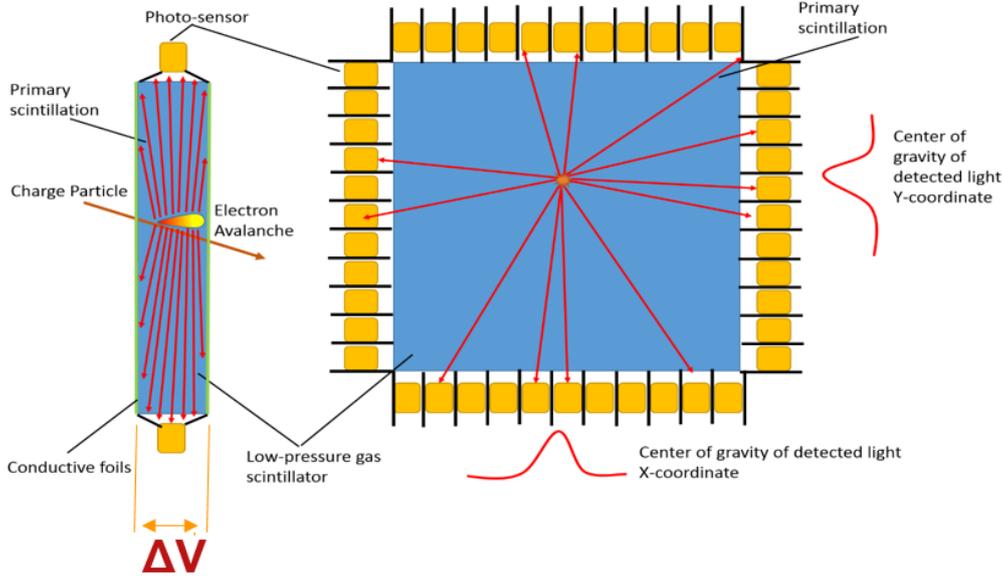

Figure 1. Operational principle of the Optical Parallel-Plate Avalanche Counter (O-PPAC).

In order to avoid large-angle and high energy straggling when the OPPAC is intended to be operated as ion transmission detector, the two parallel plate electrodes may be made of uniform, metalized polymer foils (a few μm thin). The foils can be glued onto suitable frames, which provides a compact and robust support for the whole detector assembly.

The scintillation light generated during the avalanche is reflected back and forth by the two metalized electrode foils and guided toward the photo-sensor arrays, leading to a high photo-collection efficiency. The four arrays of collimated photo-sensors are sandwiched between the two electrode frames (Figure 1) and the full assembly can be kept vacuum-tight by using rubber O-ring seals, plastic/metal gasket sealants or adhesives compounds (e.g. Hylomar). The homogeneity of the two thin metalized electrodes provides an uniform energy loss across OPPAC effective area. In addition, the total thickness of the OPPAC detector is less than a few mg/cm$^2$, ideal as a position-sensitive heavy-ions transmission detectors for tracking/timing applications.

**2.2 Choice of the scintillating gas**

The pressure and type of the scintillating gas mixture that fills the detector depend mainly on the requirement of the specific application and its experimental conditions, including low operating voltage, high charge/scintillation yield, high rate capability, and good time resolution.

Pure noble gases, such as argon (Ar) and xenon (Xe), are efficient scintillating gases. However, noble gases generally emit at very short wavelengths (e.g. Ar and Xe excimer spectra peak at 120 and 178 nm respectively), and consequently solid-state photo-sensors would require the use of wavelength-shifters (e.g. tetraphenyl butadiene - TPB) to guarantee a sufficient photon detection efficiency. High scintillation yield can be obtained from noble gases with small admixtures of impurities with smaller excitation potential, which shift the light to a more suitable (visible) wavelength range. Even in the presence of tiny amounts of certain impurities, the excimer emission will be suppressed whereas the emission of the impurities will dominate [28].

In this conditions, the primary (noble) gas acts as detector medium, providing an intense prompt scintillation light on the short wavelength range, while the impurities will serve as wavelength-shifter. Examples of gas molecules that may work as efficient wavelength shifter are triethylamine (TEA), trimethylamine (TMEA), nitrogen ($N_2$), carbon-dioxide ($CO_2$), methane ($CH_4$), etc. [29].

The tetrafluoromethane ($CF_4$) gas is another excellent electroluminesce medium, characterized by a unique combination of physical properties; it is a heavy molecule composed only of light atoms which leads to high stopping power and low gamma-sensitivity; high electron drift velocity and relative low electrons diffusion [30]. The scintillation process is fast, characterized by a short decay time - at least 90% of the light is emitted with a characteristic decay time of 15 ns or shorter [31], [32].

In one of our recent works [33], we have recently measured the correlation between secondary scintillation yield and avalanche charges in low-pressure $CF_4$ produced in a small-area (3x3 $cm^2$) PPAC detector. The light was recorded with a single Hamamatsu Multi-Pixel Photon Counter (MPPC) type VUV-3 [34], mounted in a geometrical arrangement similar to the one proposed in this work. We have found an excellent charge-light correlation and an electroluminescence yield in the range 0.01-0.15 photons/electron, depending on the reduced field applied between the PPAC electrodes and the amount of impurity present in the gas. In view of these results, the O-PPAC prototype investigated in this work was operated in $CF_4$ at 25-30 torr.

## 3. Geometry optimization methods

The conceptual design of a 10x10 $cm^2$ O-PPAC detector has been performed via a systematic Monte Carlo simulation study based on GEANT4 toolkit [24]. The parallel-plate avalanche chamber and the optical readout have been modelled based on an existing PPAC design (used at National Superconducting National Laboratory - NSCL) and on commercially-available SiPM devices. The scintillation readout comprises arrays of optically isolated SiPMs, each of them acting as one-dimensional position-sensitive photo-sensors. The simulation allows to adjust the geometrical parameters of the various detector components in order to determine an optimal configuration in terms of photon collection efficiency (signal-to-noise) and spatial resolution. The primary objectives include optimization of the effective area of the individual photo-sensor (A), the pitch (p) and the total number of photo-sensors in the array, as well as the dimension (L=length and W=width), the material and the optical properties (surface reflectivity) of the collimators (Figure 2).

It is worth pointing out that GEANT4 contains very extensive and flexible optical physics capabilities, including the possibility to generate and transport primary scintillation photons [36]. The simulation of the avalanche process is also possible by interfacing GEANT4 and Garfield++ [37], though with extremely long computing time. However, the simulation of the full Townsend avalanche process combined with the modelling of the electroluminescence mechanism is not currently possible. To overcome this limitation, the modelling of the O-PPAC operation in GEANT4 has been simplified by generating only the primary scintillation emitted along the track of the impinging charged particle, but with an effective primary scintillation yield (number of scintillation photons per deposited energy) that corresponds to the actual electroluminescence production in the course of a typical avalanche process.

For instance, alpha particles of 5.5 MeV that cross the 3 mm PPAC volume, filled with 25 torr $CF_4$, deposit an average energy of 19±5 keV (energy loss estimated by the simulations) into the PPAC gas medium, which correspond to a total of around 350 primaries. Assuming a stable PPAC operation at a gain of a few time $10^3$, the total number of avalanche electron is around $10^6$. The electroluminesce yield in $CF_4$ at low pressure, recently measured from a detailed study on light/charge correlation (see ref. [33]), reaches values of the order of 0.1 photon/electron. Thus, a typical PPAC avalanche process, under the conditions specified above, will generate a total amount of $10^5$ photons on average. Accordingly, the effective "primary" scintillation yield (number of scintillation photon per keV of energy released by the impinging charged particle) used in the GEANT4 simulation corresponds to a few time $10^3$ photons/keV (for a more conservative computation, the effective scintillation yield value was set to $2.5\times10^3$ photons/keV).

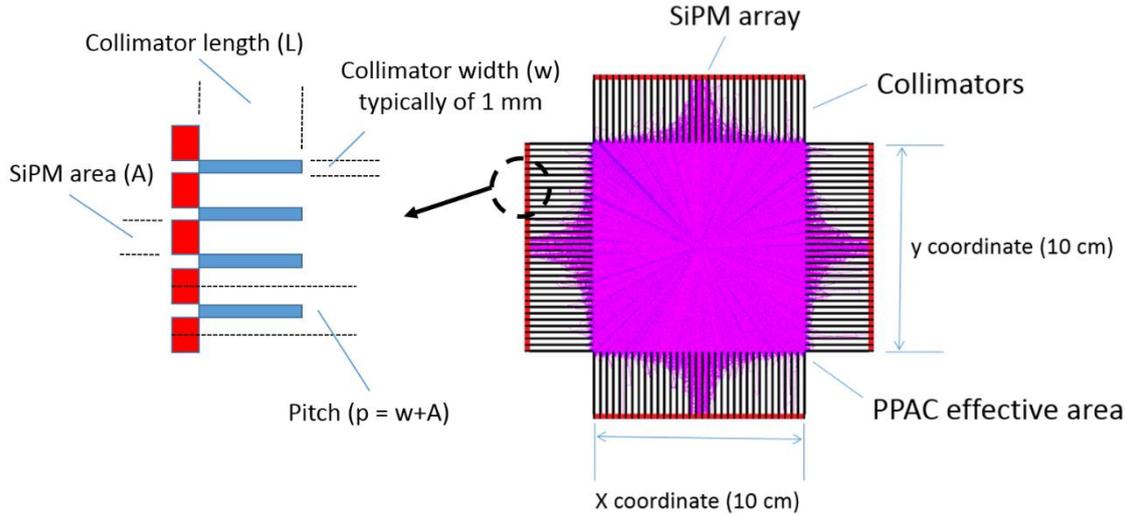

Figure 2. Snapshot of a GEANT4 simulation run: the purple lines are tracks of electroluminescence photons generated during the avalanche process, triggered by a charge particle (5.5 MeV alpha particle) impinging at the center of the O-PPAC effective area (10x10 $cm^2$). On the left side a schematic drawing of the collimator system place in front of the SiPM arrays.

Figure 2 depicts a snapshot of the simulation, showing tracks of scintillation photons (purple lines) generated during a typical avalanche process; the latter was triggered by a charge particle (5.5 MeV alpha particle) impinging at the center of the PPAC area. The figure shows also details of the collimated SiPM arrays on the four sides of the PPAC (red square). In this simulation, the PPAC volume ($3\times50\times50$ $mm^3$) was filled with $CF_4$ at 25 torr; the cathode and anode metalized (Al) foils were characterized by a photons reflectivity of 90% (dielectric-metal surface). Each array consists of 33 photo-sensors with an effective area of $2\times3$ $mm^2$ and a pitch of 3 mm. The collimators were 3 mm long and made of Teflon (dielectric-dielectric).

One of the most important factors that drives the design of the optical readout, as well as the selection of the SiPM technology, is the matching between the spectral responsivity of the optical readout technology and the secondary emission spectrum of the scintillating gas - $CF_4$ in this work. As shown in Figure 3 (red line), the secondary scintillation light in $CF_4$ consist of two broadband emissions [38]–[40]: a continuum in the UV wavelength range (200-500 nm) peaked at around 290 nm, and a visible wavelength region (500-800) peaked at 620 nm. While the broad band in the visible region is explained as the result of the excitation of a Rydberg state of the $CF_4$ molecule that dissociates into an emitting $CF_3$ fragment [41], the origin of the UV emission is explained as transitions of $CF_4^+$ excited ions [42].

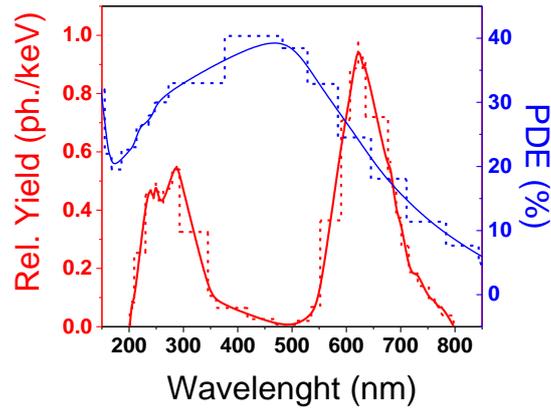

Figure 3. Emission spectrum of $CF_4$ expressed in term of relative scintillation yield (red line, data take from [38]) and the photon detection efficiency (PDE) of the Hamamatsu VUV-3 MPPC photo-sensor (blue line, data provided by [34]). The graph also shows the simplified models of the emission spectrum (red dash line) and the photon detection efficiency (blue dash line), used in the GEANT4 simulations.

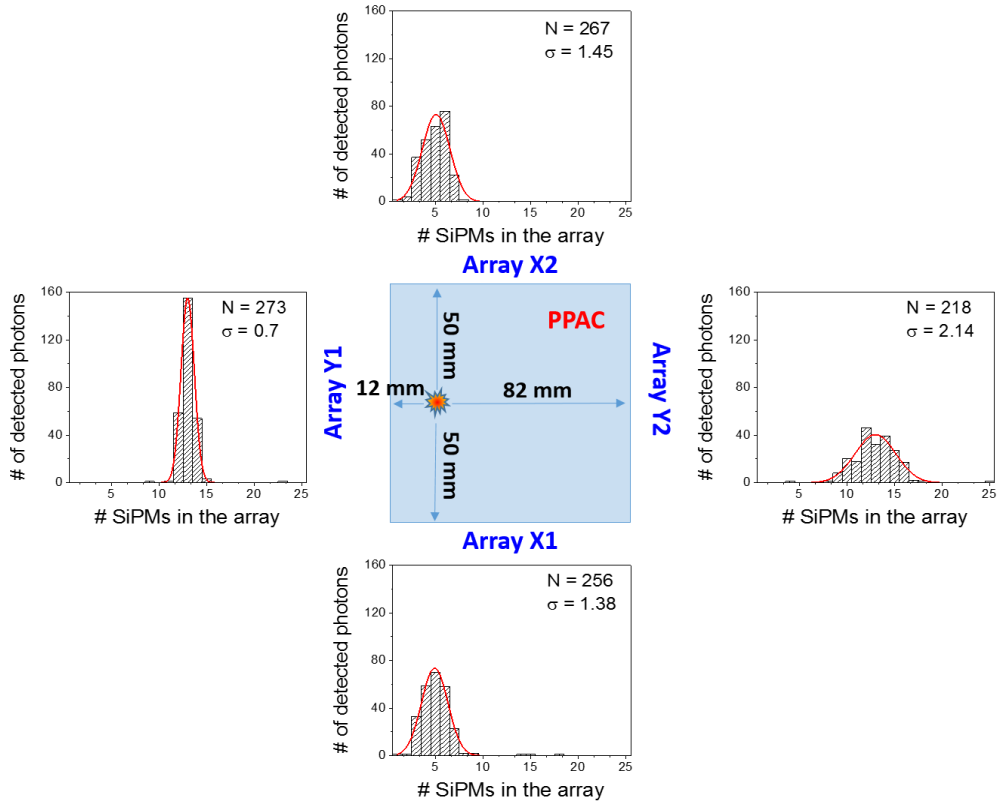

Figure 4. Example of a GEANT4 simulation output: the light spot is recorded by position-sensitive photo-sensor arrays (comprises of 25 SiPM), displaced along the four side of the PPAC. The graphs also shows the total number of photoelectrons detected by each array and the standard deviation of the Gaussian fits.

As far as the SiPM technology is concerned, the highest photo-detection efficiency may be achieved with devices characterized by a broad spectral responsivity, ideally extended to the full emission spectra of the scintillating gas. In this work we have modelled the O-PPAC optical readout based on the Hamamatsu Multi-Pixel Photon Counter (MPPC) type VUV-3 [34]; the latter has a responsivity extended to Vacuum Ultra-Violet (VUV) light, down to a wavelength below 150 nm, and it has an excellent photon-counting capability. Typical Photon Detection Efficiency (PDE) of the VUV-3 MPPC, including after-pulse and cross talk effects (neglected in the simulation), is shown in figure 3 (blue line) and is compared to the spectral emission of the $CF_4$ (red line). The optical readout of the PPAC, as well as the emission spectrum of the $CF_4$, are modelled in GEANT4 using the discretized distributions shown in Figure 3 (dash lines). The output of the simulation consist of detailed information on the spatial distribution of the photons detected by the four SiPM arrays, placed around the O-PPAC effective area (figure 4). For a more realistic modelling of the SiPM response, a conservative 10% intrinsic photon-equivalent peak resolution has also been included, together with a Poisson white noise ($\lambda$=6 photoelectrons).

The reconstruction of the position of the avalanche, which corresponds to the position of the charged-particle impinging onto the detector area, is achieved by combining the data recorded by the four SiPM arrays. This task can be performed in several ways, using various algorithms processing the full set of data readout by the photo-sensor arrays. The simpler method consists of computing the arithmetic mean of the center of gravity of the light spots recorded by the two opposing arrays; two arrays per coordinate. However, because of the finite dimension of the arrays, the light distributions of all the events occurring in the proximity of the O-PPAC border will be truncated, resulting into a distorted image response – this is generally solved by discarding the event on the border, with a significant loss of detector effective area.

A more uniform localization capability may be achieved by a dedicated algorithm that include the analysis and processing of other relevant parameters of the detected scintillation light signals. For instance, the total amount of photoelectrons detected by an array ("N" in Figure 4) depends on how distant is the avalanche from that photo-sensor array: the closer the avalanche, the larger is the amount of detected light. In the same fashion, the width ("$\sigma$" in Figure 4) of the scintillation light distribution sensed by a position-sensitive photo-sensor array is also related to the position of the avalanche.

The localization of the impinging particle along one dimension x (y) is finally computed as arithmetical mean of the two corresponding arrays' distributions means $P_{x1,x2}$ ($P_{y1,y2}$), weighted by the total number of detected photons $N_{x1,x2}$ ($N_{y1,y2}$) and the dispersion of the distribution $\sigma_{x1,x2}$ ($\sigma_{x1,x2}$), such that (similar for y coordinate):

$$x = \left( \frac{P_{x1} \cdot N_{x1}}{\sigma_{x1}} + \frac{P_{x2} \cdot N_{x2}}{\sigma_{x2}} \right) \Big/ \left( \frac{N_{x1}}{\sigma_{x1}} + \frac{N_{x2}}{\sigma_{x2}} \right) \quad \text{(eq. 1)}$$

where the means (P), the amplitudes (N) and the dispersions ($\sigma$) have been computed for the Gaussian fit of the light distribution detected by the individual photo-sensor arrays (1,2).

## 4. Design optimization results

### 4.1 Photo-sensor effective area and collimator length

The granularity of the optical readout and the dimensions of the collimators (length and width) affect the position resolution. On one hand, a larger number of photo-sensors results in better sampling of the light distribution. On the other hand, large granularity implies smaller effective area for each pixel, which leads to a smaller number of detected photons per event, and thus a worse signal-to-noise ratio (SNR). Similarly, a lengthy collimator system allows to narrow the light distribution recorded by the photo-sensor array towards its peak, which in general results in a more accurate localization of the light spot, at the expense of a lower photon statistic.

The best localization capability will be achieved as a tradeoff between effective area of the photo-sensors and the granularity of the array, as well as by compromising collimation length and SNR. The only dimension that was predefined and fixed is the width of the collimator walls (1 mm), imposed by manufacturing constrains of the SiPMs.

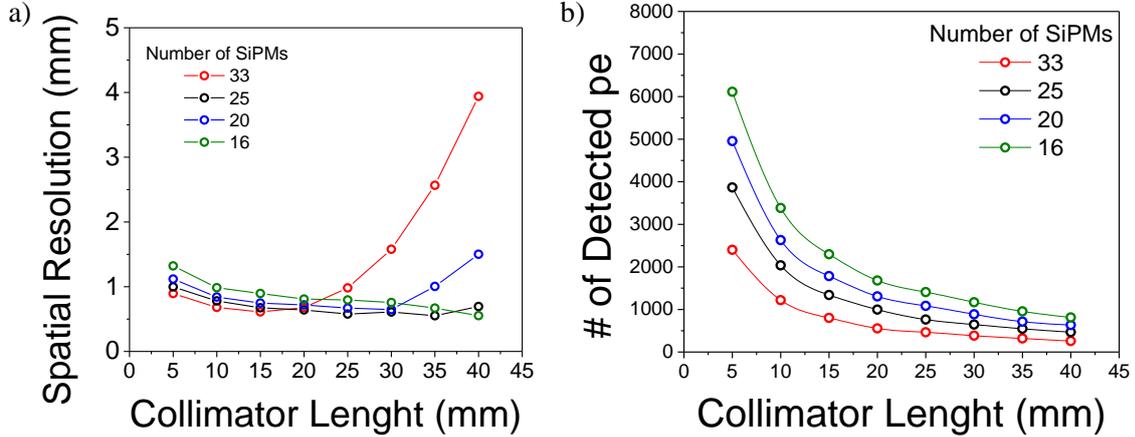

Figure 5. Position resolution (part a) and total number of detected photons as function of the collimator length, for SiPMs arrays of different granularity - the number of photo-sensor per array ranges from 12 to 33 elements.

Figure 5a summarize the results of the simulation study on the position resolution as function of the collimator length (ranging from 5 to 40 mm), for 10 cm long array with 33, 25, 20, 16 photo-sensors. The size of the area of the individual pixels range from 2x3 mm$^2$ to 6x3 mm$^2$, correspondingly. The data were obtained through a series of simulations (1000 events per each data point) of a 10x10 cm$^2$ OPPAC, filled with a 30 torr $CF_4$, traversed by 5.5 MeV alpha-particles at the center of its effective area – all the computer-simulated alpha-particles tracks were generated at the same point, so the image of the tracks on the OPPAC plane is a point-like source and the width of its image projection relates to the OPPAC spatial resolution. The energy deposited by the alpha-particle in the OPPAC gas gap is of the order of 20 keV, on average. As shown in Figure 5a, the best spatial resolution (approaching 0.5 mm FWHM) is achieved for 20-25 pixels per array and for collimator lengths of 15-20 mm; the latter is the most cost-effective optical readout configuration.

**4.2 Linearity and image quality**

The spatial variations of photon collection efficiency, due to the hard collimation, and the modulation of the optical signal, due to the finite dimension of the photo-sensor, are possible sources of image distortion and loss of linearity across the OPPAC effective area.

The image linearity and the homogeneity of the imaging system response were assessed by processing the Gaussian fit of the image projections of some computer-simulated (GEANT4) alpha-particles pencil-beams (Figure 6a), evenly distributed along one coordinate (x-axis) at a distance from one another of 8 mm. In particular, the optimum geometrical configuration, comprising an optical readout with 25 photo-sensors and 25 mm long collimator walls, has been investigated and evaluated.

The effects induced on the image by distortions can be characterized by the Integral Non Linearity (INL), defined as the deviation of the measured image holes centroid from their real locations on the OPPAC plane. Geometrical distortions represent a deviation from rectilinear projection.

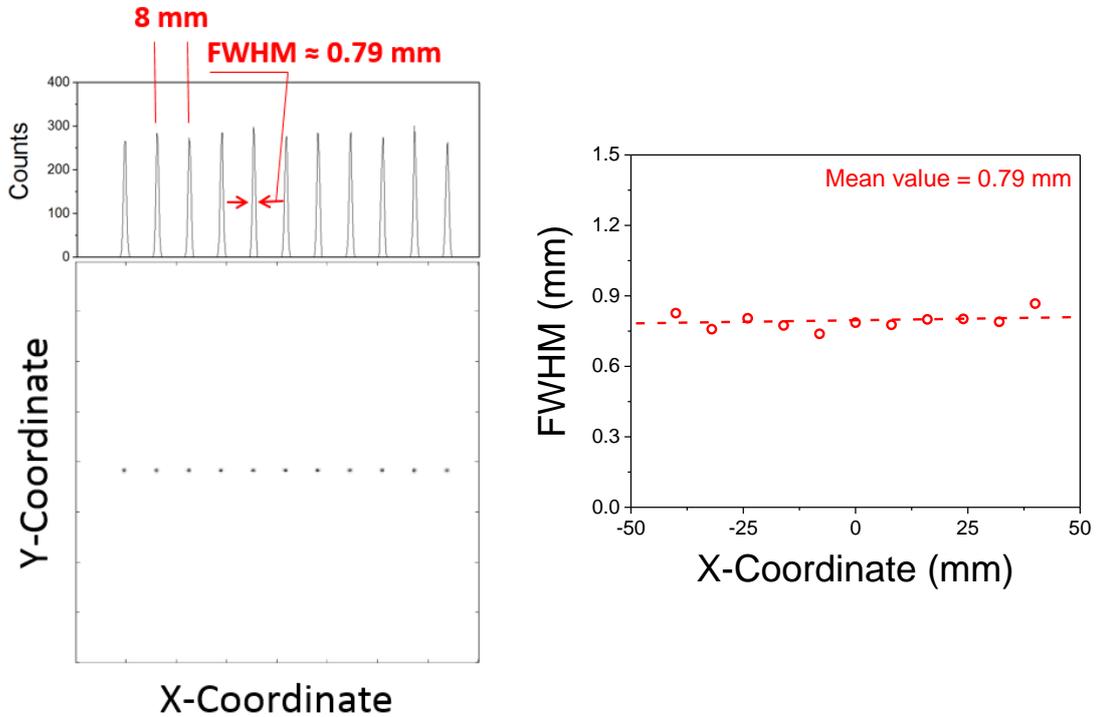

Figure 6. Part a: simulated OPPAC images of several point sources, disposed along the x–coordinate of the OPPAC effective area. The image profile and extrapolated spatial resolution is also shown at the top of the image. Part b: Uniformity of the spatial resolution (FWHM) along the OPPAC length (x-coordinate).

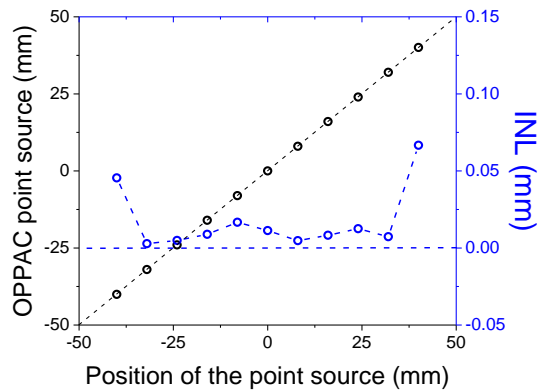

Figure 7. Integral Non Linearity (INL) of the OPPAC detector (25 sensors per array with 25 mm long collimators), calculated from the centroid of peaks in the projected 1D images shown in Figure 6a.

As shown in Figure 6b, there is not significant variation of the spatial resolution along the OPPAC, being of around 0.8 mm (FWHM) except for slightly increase at the edge of the effective area, with variations below 10%. However, boarder effects became more evident in term of INL. Distortion may also arise from several other effects, such as mechanical defects of the collimator system, variation of gas gain across the OPPAC effective area, preamplifiers cross talk, and different responses of the front-end electronics channels used to process the SiPMs' signals. A digital correction algorithm may be further implemented to compensate systematic geometry distortions or inhomogeneity defects.

## 5. Performance evaluation of a 1D OPPAC prototype

A first feasibility study on the O-PPAC concept was carried out by investigating the properties of a 10x10 cm$^2$ detector prototype (Figure 8a), equipped with a single 1D photo-sensor array. The latter consists of 10 Hamamatsu VUV-3 MPPCs, commonly biased at an operational voltage of 56 Volt through a 10 kΩ resistors chain. Although the MPPCs have an effective area of 3x3 mm$^2$, their ceramic package has a total width of 6 mm, which limits the granularity of the optical readout and constrains the performance of the O-PPAC prototype in terms of localization capability. Nevertheless, the experimental investigation of the OPPAC prototype was extremely important for validating the Monte Carlo simulation package described in section 4. It allows to characterize the response of the optical readout to electroluminescence light in a parallel-plate configuration, in terms of photon collection efficiency and spatial uniformity. Moreover, it also permits an overall evaluation of materials and fabrication procedures.

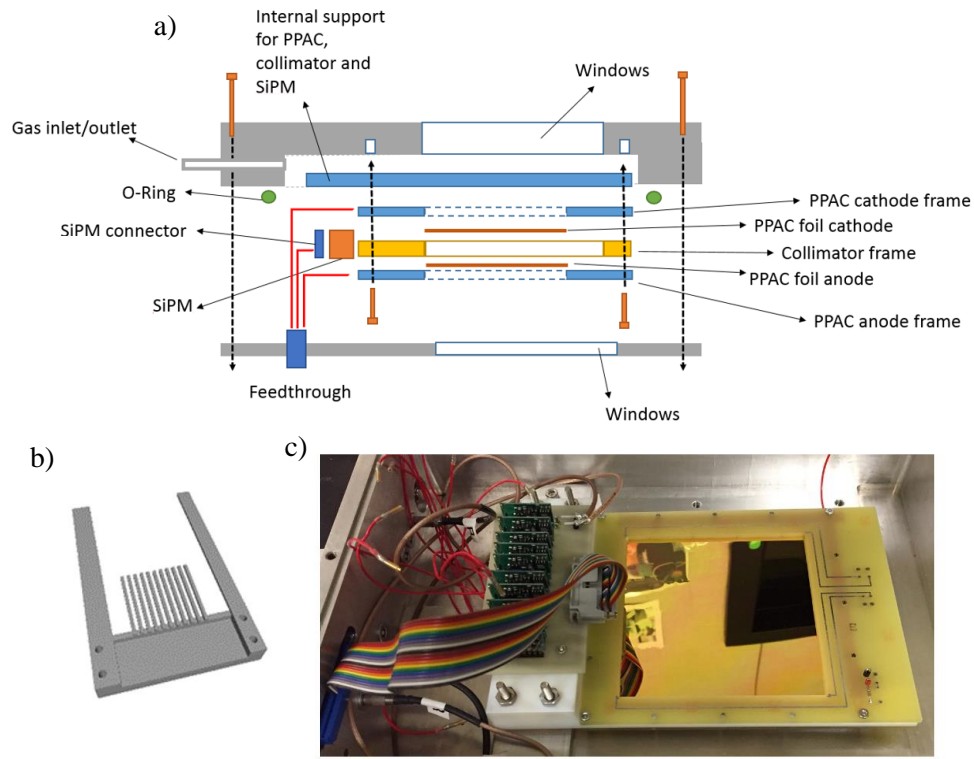

Figure 8. Part a: schematic drawing of the mechanical assembling and components of the O-PPAC prototype. Part b: 3D rendering of the collimator structure that holds the SiPM arrays. Part c: photograph of the O-PPAC prototype.

The light signals of each individual MPPC was processed through custom-made, general purpose charge integrating pre-amplifiers (model SR CHARGE8VLNDC [43]), mounted inside the detector vessel. The output of the preamplifiers were fed to a multi-channel shaping amplifier (Mesytec model MSCF-16) and the resulting distribution of pulse-heights, sensed by the MPPCs array, was then recorded by a multi-event Peak Sensing ADC (CAEN model V785). The data were further stored and processed by a custom-made software. The timestamps for the event triggers of the ADC unit were either provided by the digitalized PPAC charge signals or by the multiplicity trigger of the MSCF-16 shaping amplifier.

The MPPCs were mounted on a PCB board and inserted in the slots of a collimator system comprises of 20 mm long slabs. Each slab was 1 mm wide and 3 mm in height – the slabs were inserted in the amplification gap between the anode and the cathode foils. The collimator system, made of Acrylonitrile butadiene styrene (ABS), was manufactured by a high-precision 3D printing machine (Objet Connex350 Multi Material 3D Printing System). The reflectance of the ABS-made collimator walls is believed to be lower than 5-10%, over the emission spectrum of the $CF_4$. Further lower reflectivity can be achieved by a suitable optical coating and will be considered for future developments.

The anode and cathode foils consisted of stretched polypropylene foils (a fraction of micron thick), with a 150 nm thin Au layers evaporated on one side. The foils were glued on a FR-4 frames, which holds HV connectors for DC voltage biasing. Details of the O-PPAC assembling, the mechanical design and a photograph of the detector vessel are depicted in Figure 8. The detector was mounted in an aluminum vessel equipped with gas outlet/inlet, HV and signal feedthroughs, and two thin (a few µm) pressure windows made of 1.5 micron aluminized polyester foils.

The O-PPAC prototype were tested in three different operational conditions:
1) Direct illumination of the MPPCs array with a low-intensity, collimated LED light source, while the PPAC was not in operation.
2) Irradiation with 5.5 MeV alpha-particles from a 241-Am source. The PPAC detector was operated in $CF_4$ at a pressure of 30 torr.
3) Irradiation with 60 A MeV $^{10,12}C$ beams, performed at the Research Center for Nuclear Physic (Japan). The PPAC detector was operated in $CF_4$ at a pressure of 25 torr.

## 5.1 Measurement with LED light

The direct measurements of the light pulse from a LED source, with no operation of the PPAC, were used to develop both the DAQ system and the software for data processing. The measurements also allow to extract some information related to the intrinsic working operation of the MCCPs arrays, used as input parameter for the layout of the simulation package. Figure 9a depicts a schematic drawing of the experimental setup: the MCCPs were directly illuminated by a low-intensity, "yellow" LED light source (model COM-095094), placed in front of the photo-sensors array. The detector vessel was light-sealed to reduced background, and kept in air at atmospheric pressure; no bias was applied to the PPAC. A small collimator, with diameter and the length of 1 mm and 5 mm respectively, was placed in front of the LED; this results in an illumination profile, projected on the MPPC arrays surface, of around 25 mm. A typical pulse-height distribution recorded by the MPPCs array, from a single light pulse (a few hundred nanosecond long), is shown in Figure 9b. The data of the pulse-height profile along the MPPCs array was fitted with a Gaussian function model, and the resulting Gaussian median point was used to localize the light-spot source.

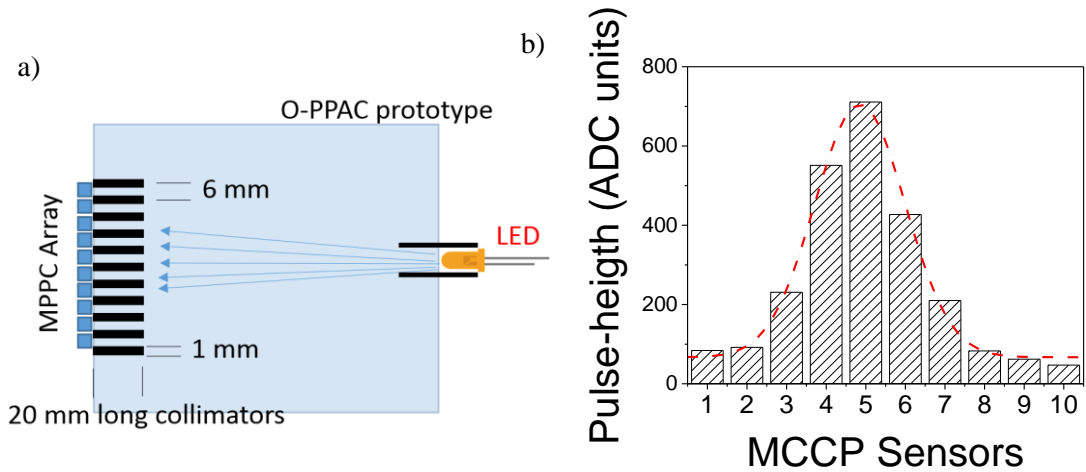

Figure 9. Part a: schematic drawing of the detector setup and LED irradiation, used for the development of the DAQ and data processing. Part b: typical pulse-height distribution (and Gaussian fit – in red) recorded by the MCCPs array from a single LED light pulse.

Figure 10 depict the distribution of the light spot position measurements (part a), as well as the pulse-height spectrum (part b) sensed by of one of the MPPC sensor (sensor number 6 along the array). The results were extracted from the analysis of 10k events. Assuming the actual size of the light spot to be negligible, the intrinsic spatial resolution (Figure 10a) provided by the optical readout resulted to be of the order of 0.64 mm. The characteristic pulse-height resolution (Figure 10b) of the MPPCs was of 11.5%, assuming a negligible event-by-event variation of the LED light intensity. The measured value of the intrinsic MPPC energy resolution was used in the Monte Carlo simulation as photoelectron peak resolution, to provide a more realistic modelling of the optical-readout response (see section 3).

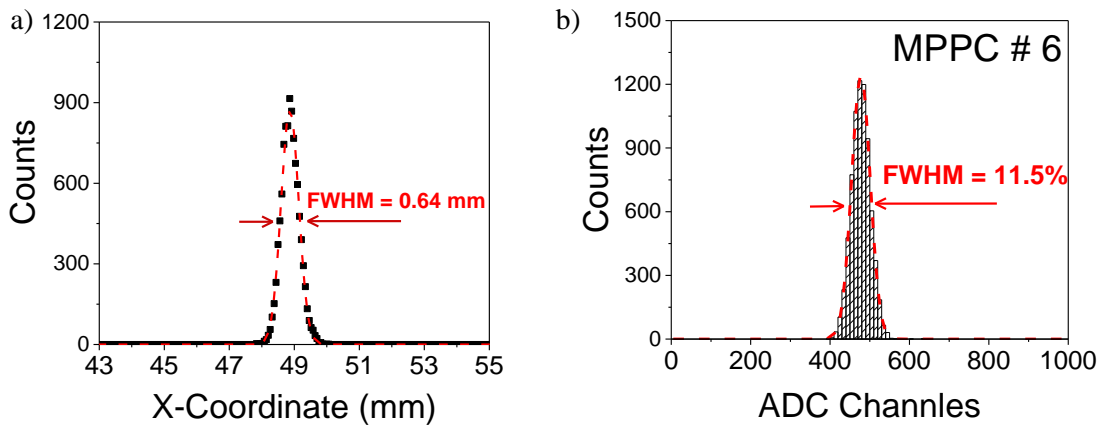

Figure 10. LED light measurements (10k events). Part a: distribution of the position of the light spot computed along the arrays coordinate. Part b: pulse-height spectrum recorded by the MPPC number 6.

**5.2 Measurements with 5.5 MeV alpha-particles**

Figure 11a illustrates the detector setup used for the measurement with the 5.5 MeV alpha-particles emitted from a low-intensity 241-Am source. To minimize the energy loss of the alpha-particles before they enter the detector effective area, the O-PPAC chamber, filled with $CF_4$ at 30 torr, contained in a larger vacuum vessel with a pressure below 10$^{-5}$ torr. The 241-Am source was placed in the vacuum vessel in front of the O-PPAC at a distance of 50 cm from the O-PPAC entrance window. The voltage difference across the PPAC gap was kept at 900 volts, which corresponds to an amplification gas gain of a few tens 10$^3$. Figure 11b shows a typical pulse-height profile of the avalanche-induced light, recorded by the photo-sensor array; the avalanche process was triggered by an alpha-particle that crossed the O-PPAC at the center of the effective area. The profile of the detected light is narrow thanks to the hard collimation, with only few MPPCs being illuminated. Figure 11b also shows the Gaussian fit used for particle localization (red dash line).

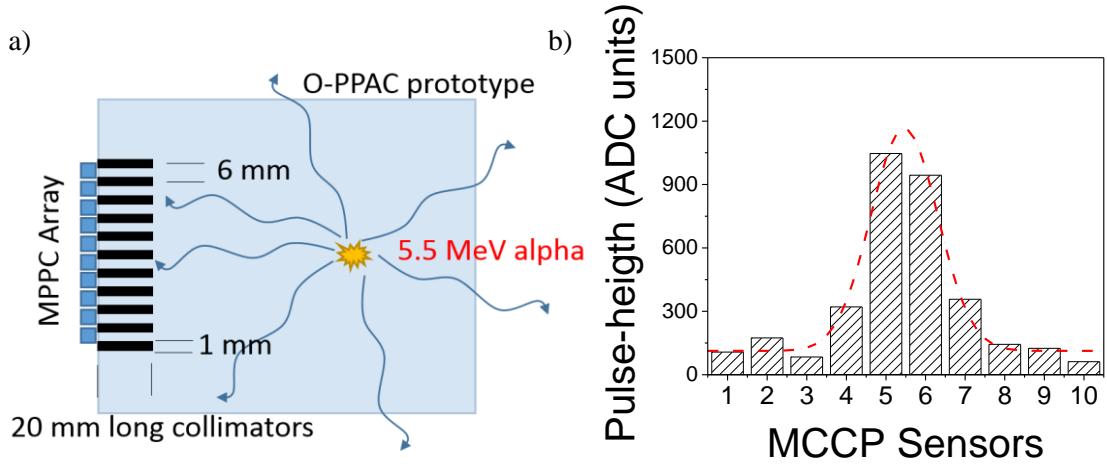

Figure 11. Part a: schematic drawing of the detector setup used for 5.5 MeV alpha-particle irradiation. Part b: typical pulse-height distribution (and Gaussian fit – in red) recorded by the MCCPs array from a single avalanche process.

The processing of the raw data was performed using Matlab [44] and the Matlab curve fitting toolbox programs. Because the Gaussian fitting approach is very inefficient for fitting truncated normal distributions, all the events having a maximum of the pulse-height distribution at the very edge of the array were discarded. This reduces the prototype localization capability to a 1D image projection with effective length of 55 mm.

Note that, while all the sensors were biased from a common HV power supply (56 Volt) using a dedicated voltage distribution circuit, the actual optimal operational voltage of MPPCs may vary within a certain range (typically a few hundred of mV). This causes variations of internal gains and responsivities of the photo-sensors placed along the array, which may result in a sizable residual inhomogeneity.

Figure 12a shows the count density profile of a uniform "empty field" irradiation with 5.5 MeV alpha-particles. The resulting image is affected by large variations that can only be corrected either by adjusting the operational voltage of each individual sensors, or by an appropriate software calibration that will take into account the variation of the photon detection efficiency of the photo-sensors.

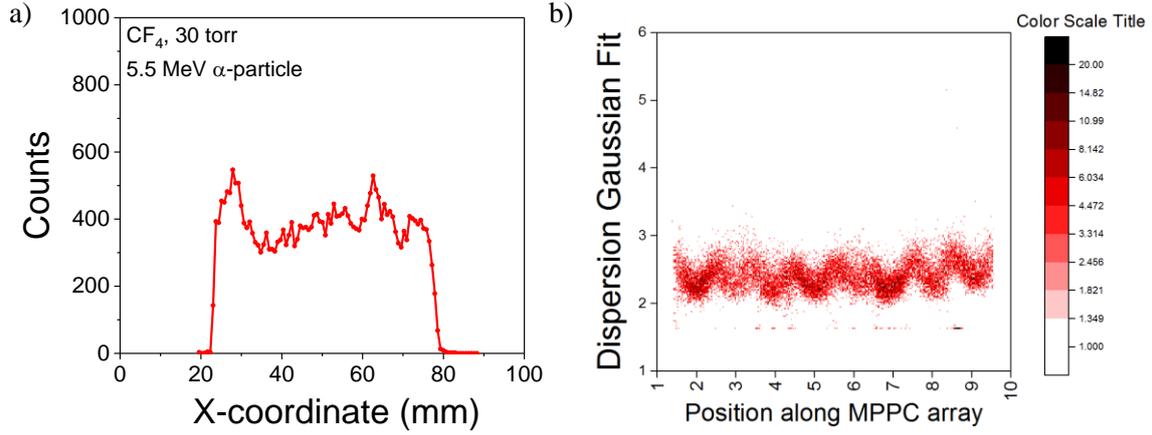

Figure 12. Part a: Raw data projection of an empty-field irradiation along the MPPC arrays. Part b: density map of the standard deviation computed from the Gaussian fits of the MPPC array pulse-height distribution.

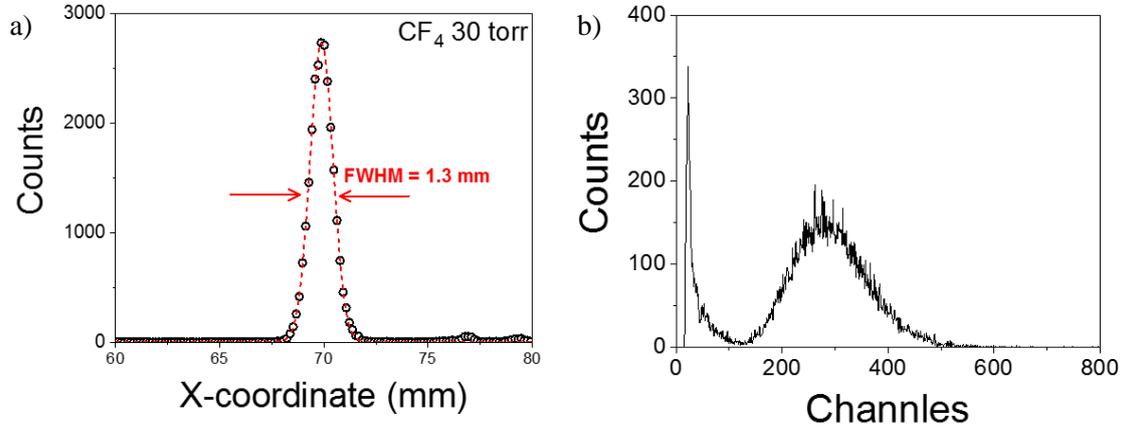

Figure 13. Part a: one-dimensional image profile of a pin-hole mask. Part b: Pulse-height spectrum of the electroluminesce light recorded by a MPPC (sensor number 6); the avalanche were triggered by 5.5 MeV alpha-particles, recorded by a MPPC.

Figure 13a depicts the image profile of a (5 mm) brass mask with a 1 mm diameter pinhole, at the center of the OPPAC effective area. The pinhole mask was deployed on top of the PPAC frame and the detector was irradiated by 5.5 MeV alpha-particles from the non-collimated 241-Am source, placed at a distance of 50 cm from the OPPAC. The measured FWHM of the pinhole profile is of order of 1.3 mm (Figure 13), so that the in an intrinsic spatial resolution resulted to be of 0.84 mm (after removing the contribution of the pinhole size); this is comparable to the value measured with the LED light source (0.63 mm). The effect of the coarse optical readout granularity (6 mm pitch) is clearly evident for the density map of the pulse-height dispersion (sigma of the Gaussian fit) and the location of the corresponding avalanche. The oscillation of the localization accuracy (sigma) along the arrays is originated by the finite dimension of the photo-sensor pixel. The modulation effects on the optical readout response increases progressively as the light spot approaches the array and depends on the granularity. Correction for the modulation effect is achieved by processing the optical readout data of the four array following the recipe described in section 3. Figure 13b illustrates a typical pulse-height spectrum of the electroluminesce light recorded by a MPPC. Similar pulse-height distribution can be obtained from the PPAC charge signals, due to the excellent light-charge correlation.

## 5.3 Measurements with $^{10,12}$C beam

The performance of the OPPAC prototype has been investigated for the first time under heavy-ion irradiation at the Research Center for Nuclear Physics (RCNP), Osaka (Japan). The detector was installed at the F2 focal plane of the Exotic Nuclear beam line (EN course) [45], and irradiated with a $^{10}$C secondary beam at 60 A MeV. The O-PPAC was filled with $CF_4$ at a pressure of 25 torr. A bias of approximately 900 Volt were typical applied across the PPAC avalanche gap, which corresponds to a gas gain approaching a value of $10^4$. Differently from the experimental setups described above, the collimators lengths were of only 5 mm (Figure 14a) in order to achieve the highest photon collection efficiency. An example of a characteristic pulse-height profile measured in this conditions is shown in Figure 14b, and it may be compared to the data shown in figure 9b. The portion of the MPPC array illuminated by the avalanche light is much larger compared to the harder collimation used previously (Figure 11b), which in reality led to a significant loss of spatial resolution despite the excellent photon counting statistics.

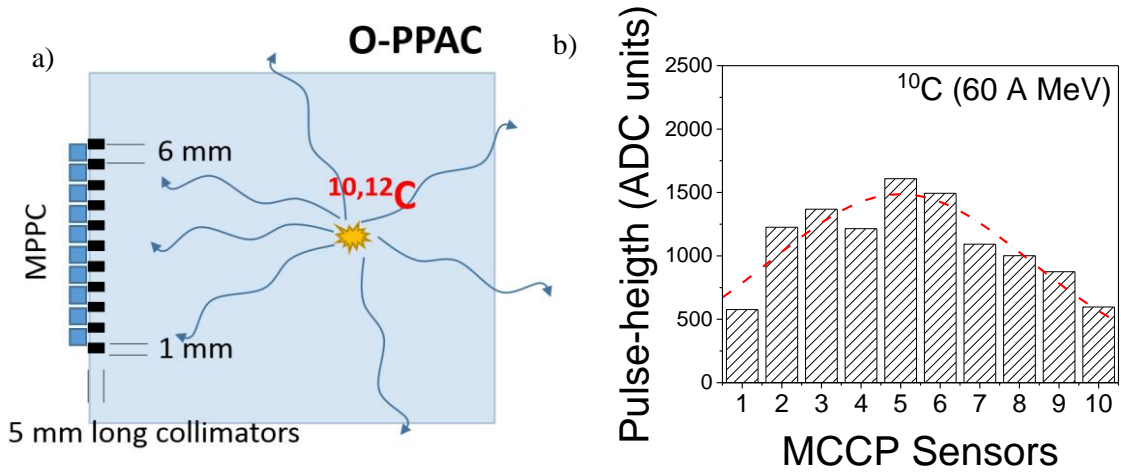

Figure 14. Part a: schematic drawing of the detector setup used at RCNP for irradiation with the $^{10,12}$C beam. Part b: typical pulse-height distribution (and Gaussian fit – in red) recorded by the MCCPs array from a single avalanche process.

Two delay-line PPACs [46] upstream the O-PPAC track the beam particles on an event-by-event basis. Note that, the two PPACs were separated one another by 480 mm and from the downstream O-PPAC by more than 1.8 meter, and they are characterized by a typical spatial resolution of around 0.5 mm (sigma), which in turn results in a large uncertainty on the localization of the beam particle on the O-PPAC plane of the order of 3 mm. As the beam was focused on the PPAC$_2$ plane and diverged downstream, the beam profile on the O-PPAC plane was larger compared to portion of area seen by the O-PPAC readout. In addition, the O-PPAC was slightly misaligned with respect to the beam axis, and therefore some further processing was necessary to extract data useful for evaluating the O-PPAC performance: only events corresponding to beam particles hitting the portion of the O-PPAC area directly seen by the optical readout were analyzed, while all the other events rejected. The beam images measured with the two PPACs, as well as the expected beam profile projected onto the O-PPAC plane, are shown in Figure 15.

Figure 16a shows the correlation map between expected positions of the $^{10}$C beam particle position measured with the O-PPAC (ordinate) and the one estimated by the delay-line PPACs tracking system (abscissa), for the process of more than 100k events. The projected/O-PPAC position measurement correlation is clearly evident, though exhibits a quite large variance due to the lack of tracking resolution. The small deviation from the linear correlation visible at one end of the edge of the optical readout range is due to the O-PPAC misalignment, mentioned before. A typical scintillation light spectrum measured by one of the MPPC in the array, under irradiation with the $^{10}$C beam, is depicted in Figure 16b. The light spectrum, similar in shape to the pulse-height spectrum measured from the PPAC charge signal, has the distinctive form a Polya distribution, which is typical for fast beam particle that originates a small number of primaries in the avalanche gap. Nevertheless, the light-spectrum spans over one order of magnitude thanks to a low limit of detection and dynamic range of the O-PPAC optical readout.

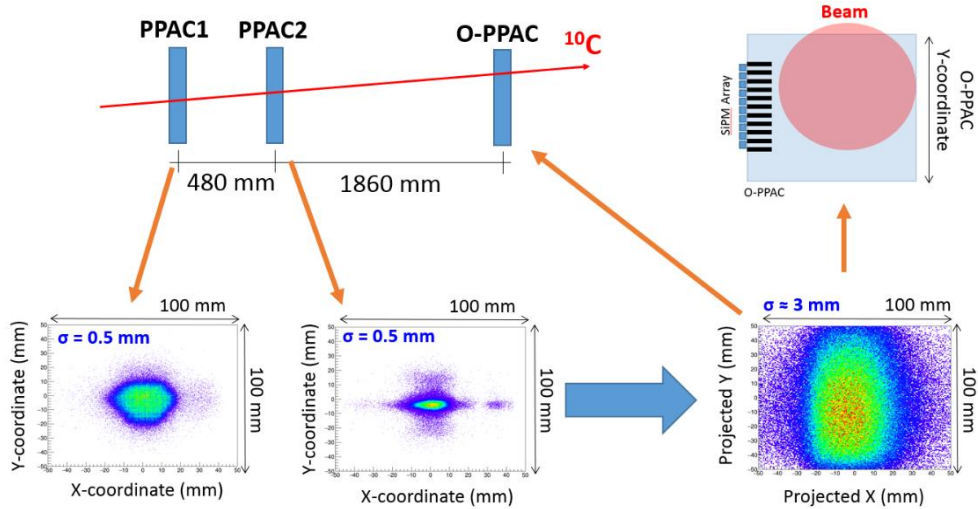

Figure 15. Schematic representation of the measurement setup, which includes the O-PPAC prototype preceded by two delay-line PPAC for particle tracking. The figure also shows the image of the $^{10}$C beam measured by two PPACs and the expected projection of the beam onto the O-PPAC surface.

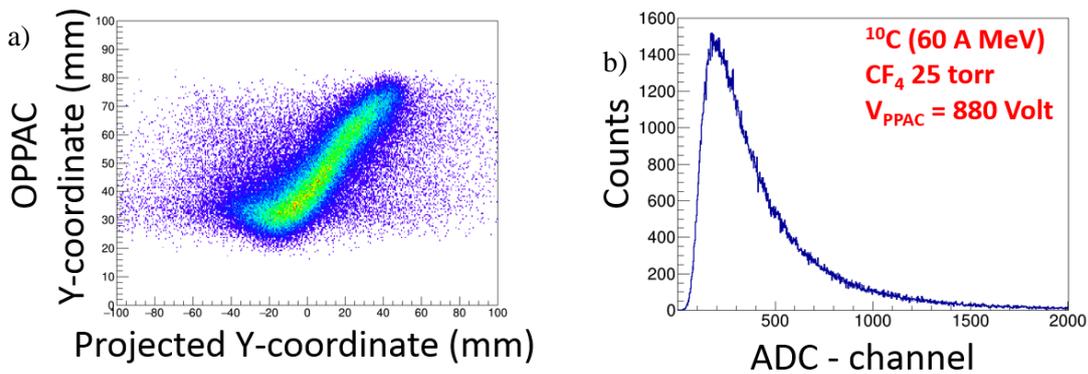

Figure 16. Part a: Correlation map obtained by plotting the expected beam particle position measured with the O-PPAC (ordinate) and the one estimated by the delay-line PPACs tracking system (abscissa). Part b: scintillation light spectrum from $^{10}$C-induced avalanche, measured with one of the MPPC in the array.

To extrapolate the spatial resolution of the O-PPAC detector, a small fraction of data corresponding to beam particles hitting the center of the OPPAC effective area (within a windows of 2 mm) was selected (Figure 17a). The profile of the corresponding position measured with the OPPAC is shown in Figure 17b. Note that, in this specific experimental conditions, the uncertainty $\delta_{measured}$ (3,91 mm) in the position measurement is actually the sum of three distinct terms:

$$\delta_{measured} = \delta_{OPPAC} + \delta_{window} + \delta_{tracking} \quad (eq.\ 2)$$

where $\delta_{OPPAC}$ is the intrinsic spatial resolution of the O-PPAC, $\delta_{window}$ is the width of the area from which the analyzed data have been selected (2 mm), and $\delta_{tracking}$ is the uncertainty of the tracking system (3 mm). From eq. 2 it is then therefore possible to extrapolate the intrinsic spatial resolution of the O-PPAC (with 5 mm length collimator), resulting of the order of 1.5 mm (sigma). This value is significantly higher compared to the one estimated early (below 1 mm FWHM). However, the discrepancy may be clearly understood in term of counting statistics and the soft photo-sensor collimation. On one hand, short collimator walls allows to collect larger fraction of scintillation photons for a higher SNR. On the other hand, a soft collimation results in a broad pulse-height distribution along the MPPC array, leading to a significant loss of accuracy in reconstructing the position of the light source. In addition, most of the avalanche has actually a low pulse-height amplitude and characterized by the typical Polya-like pulse-height distribution, with a generally low counting statistics. All these factors contributed to a general degradation of the position resolution, which suggest that higher gas pressure, a better collimation and a higher optical readout granularity are mandatory in the case of fast beam.

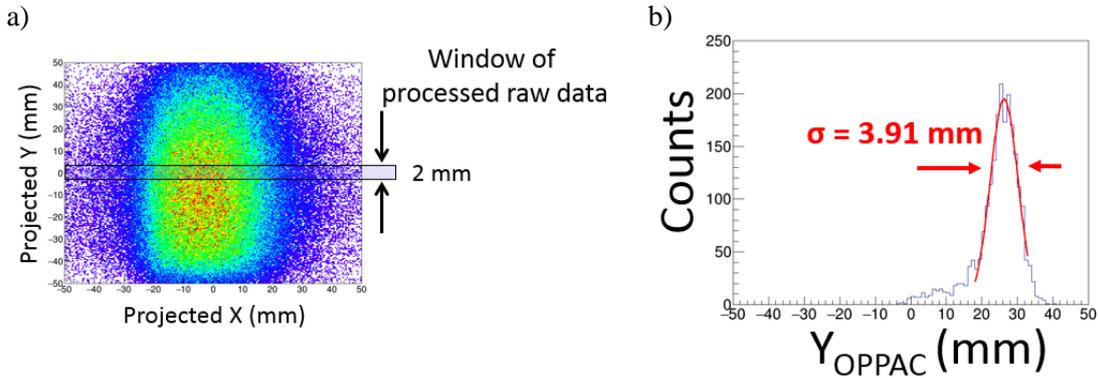

Figure 17. Part a: Windows of beam particle events (2 mm wide) selected for the computation of the OPPAC position resolution. Part b: profile of the image of the selected event measured with the OPPAC.

## 6. Summary and conclusion

The operation mechanism and properties of a novel gaseous detector concept for heavy-ion tracking and imaging, the Optical parallel-Plate Avalanche Counter (O-PPAC), has been presented and discussed for the first time. The detector consists of a parallel-plate avalanche counter filled with high-yield scintillating gas (i.e. CF4) and equipped with an optical readout. The latter is based on arrays of small-area, collimated photo-sensors (SiPMs or APDs) or position-sensitive vacuum/gas photomultipliers, placed around the four edges of the avalanche gas gap area. The position of the impinging charged particle is derived from the analysis of the light distribution sensed by the optical readout (e.g. form the center of gravity of the recorded scintillation light).

The O-PPAC offers has several noteworthy operational advantages, including a fast pulse propagation (good time resolution); insensitivity to electronic noise or RF pick-up problems because the readout is decoupled from the electron avalanche volume; large dynamic range with respect to the particle's mass and energy due to the extremely low limit of detection of the optical readout (down to single photoelectrons); good detection efficiency due to a large light yield; more uniform response and less angular/energy straggling due to a lower material budget.

A systematic Monte Carlo simulation study, focused on O-PPAC geometry optimization and overall detector performance evaluation, have been presented. It was found that the optimal detector assembly (20-25 photo-sensor over 10 cm long array, equipped with 15-20 mm long collimators walls) provides a sub-millimeter localization capability, with good homogeneity across the full detector active area.

We present the first experimental results with a test-bench detector system, comprising a 10x10 cm$^2$ PPAC avalanche area optically readout by a single MPPC array. The arrays are comprised of 10 MPPC (type VUV-3), with a pixel pitch of 6 mm. The detector was tested under different irradiation conditions, providing a submillimeter spatial resolution.

Applications of the O-PPAC include tracking, imaging and time-of-flight measurements for the heavy-ions experiments, in particularly for physics at the new generation of rare-isotope beam facilities, as alternative to conventional PPAC or drift chamber detector. Other potential applications include position-sensitive transmission imaging for beam diagnostics in hadron-therapy, heavy-ions range radiography, and scintillation counter as Compton scattering gamma ray imaging camera.

## Acknowledgments

The authors would like to thank the RCNP laboratory for providing the $^{10}$C and $^{12}$C beams used for to test the OPPAC prototype. H.J. Ong and T. Furuno, RCNP, are greatly acknowledged for their support during the experiment.